\documentclass[fleqn,usenatbib]{mnras}
\usepackage[utf8]{inputenc}

\title{On the Orbital Velocity of Isolated Galaxy Pairs:  
a test of gravity in the low acceleration regime. }

\author[R. Scarpa et al.]{Riccardo Scarpa,$^{1,2}$\thanks{E-mail:
riccardo.scarpa@gtc.iac.es}
Renato Falomo,$^{3}$
Aldo Treves$^{4,5}$
\\
$^1$ Instituto de Astrofisica de Canarias, C/O Via Lactea, s/n, E-38205 La Laguna, Tenerife, Spain\\
$^2$ Departamento de Astrofsica, Universidad de La Laguna, E-38206 La Laguna, Tenerife, Spain\\
$^3$INAF - Osservatorio Astronomico di Padova, vicolo dell’Osservatorio 5, I-35122 Padova, Italy\\
$^4$ Universit\`a dell’Insubria, via Valleggio 11, I-22100 Como, Italy\\
$^5$ INAF - Osservatorio Astronomico di Brera, via Bianchi 46, I-23807 Merate (Lecco), Italy\\
}

\date{Last Rev. 6 December  2021}

\usepackage{natbib}
\usepackage{graphicx}

\begin{document}
\label{firstpage}
\maketitle

\begin{abstract}
The dynamics of isolated galaxy pairs represents an important tool to investigate the behavior of gravity in the low acceleration regime. Statistical analysis of a large sample of galaxy pairs  led to the noticeable discovery of a region of preferred 3-dimensional velocities centered at $\sim 150$ km/s and $\sim 100$ km/s wide, a feature hard to justify in the context of numerical simulations of cosmological structure formation. 
It is shown here that such a feature is expected within the framework of the Modified Newtonian Dynamics, which, however, predicts it to be centered at $\sim$ 170 Km/s.
\end{abstract}


\section{Introduction}
Many lines of investigations in present day astrophysics support the existence of large amounts of non-baryonic dark matter (DM hereafter) in galaxies and in the Universe at large. The most obvious example being the asymptotically flat rotation curve of galaxies, which posits for the existence of massive halos of DM. 
Considerable fine tuning of DM content in galaxies is however required to justify their observed properties.  The most striking example is possibly the baryonic Tully-Fisher relation (see e.g. McGaugh 2012 and reference therein). Because of this, over the years various  proposals have been made to find alternative explanations, not involving DM. 
In particular, it has been shown that a specific  modification of Newtonian dynamics (MOND, Milgrom 1983a,b,c) is able to describe many, if not all behaviours of galaxies generally ascribed to the presence of DM.

In search for new tests for both DM and alternative gravity theories, we focus here on isolated galaxy pairs as {\bf potential} important probes of dynamics under gravity. In particular we examine in this paper the newly published Isolated Galactic Pair Catalogue (Nottale and Chamaraux 2018, NC18a hereafter). We consider the detailed discussion presented in Nottale and Chamaraux (2020, NC20 hereafter), which starting from the available observable parameters -- radial velocity difference, projected separation, and luminosity of the two components of each pair -- were able to statistically recover the 3-dimensional orbital inter-velocity distribution. This led to the remarkable discovery of a preferred region for the orbital velocity ($\sim$ 150 km/s). We discuss here the difficulties of explaining such a preferred orbital velocity within a conventional Newtonian framework, and consider alternative dynamical scenarios.

\section{ Isolated Galaxy Pairs Catalogue}
\label{Sextcat}
The Isolated Galaxy Pairs Catalogue (IGPC) was constructed by NC18a extracting the pairs from the HyperLEDA  extragalactic database (Makarov 2014).  The following selection criteria were used: 
1) absolute B magnitude of galaxies M$\leq -18.5$; 
2) projected separation $r_p< 1$ Mpc; 
3) radial velocity $3000<v_r<16000$ km/s;
4) radial velocity difference (inter-velocity) $|\Delta V|< 500$ km/s;
5) exclusion of multiplets, each member is the closest to the other one; 
6) isolation criterion $\rho=r_3/r_p$, where $r_3$ is the projected distance of the next nearest galaxy to the pair with $|\Delta V|<500$ km/s. Taking $\rho \geq 2.5$ the catalogue contains N=13092 pairs satisfying the previous requirements, N=7449 for $\rho \geq 5$,  and  N=4599 for $\rho \geq 9$.

\begin{figure}
\hspace*{-0.65cm}
\includegraphics[scale=0.48]{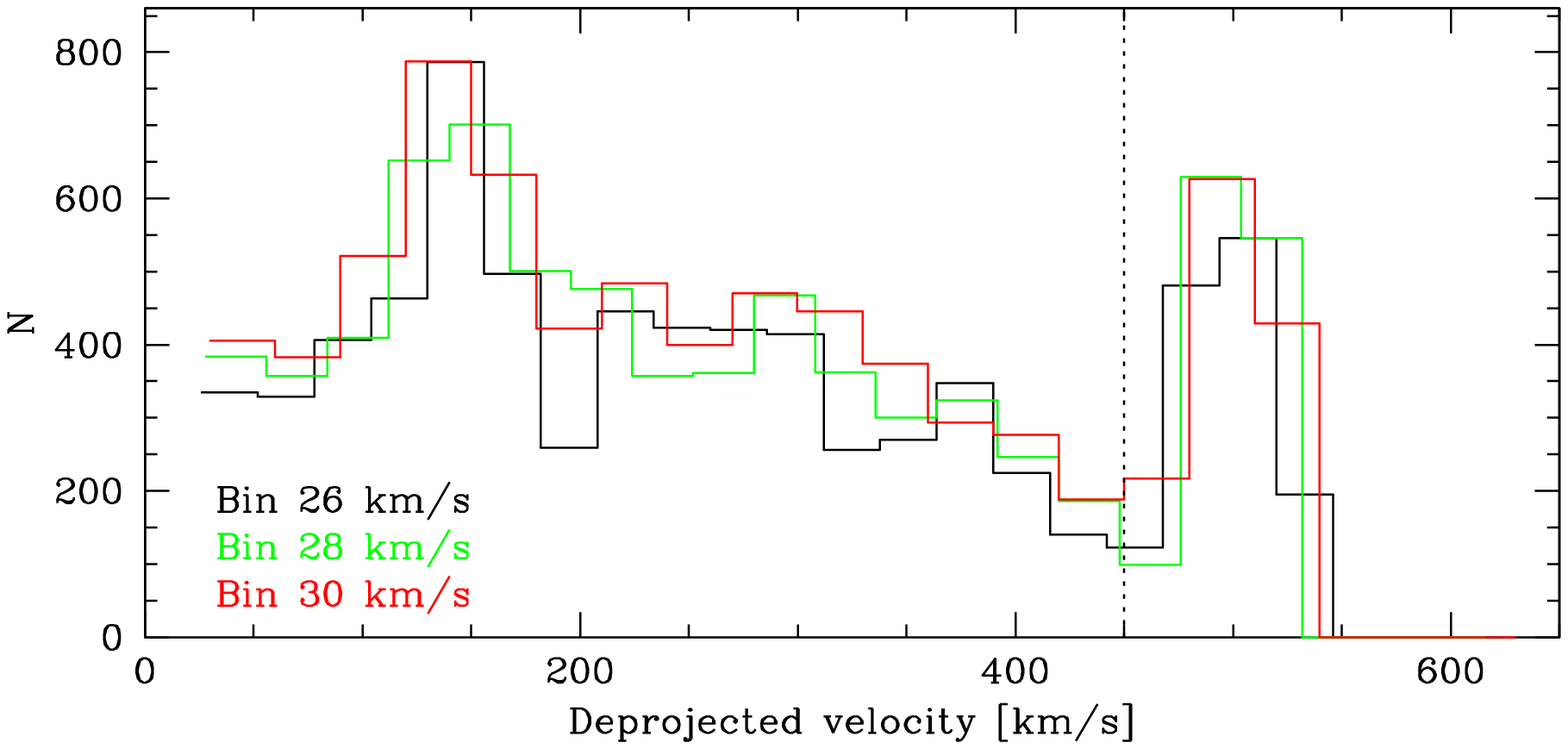}
\hspace*{-0.65cm}
\includegraphics[scale=0.48]{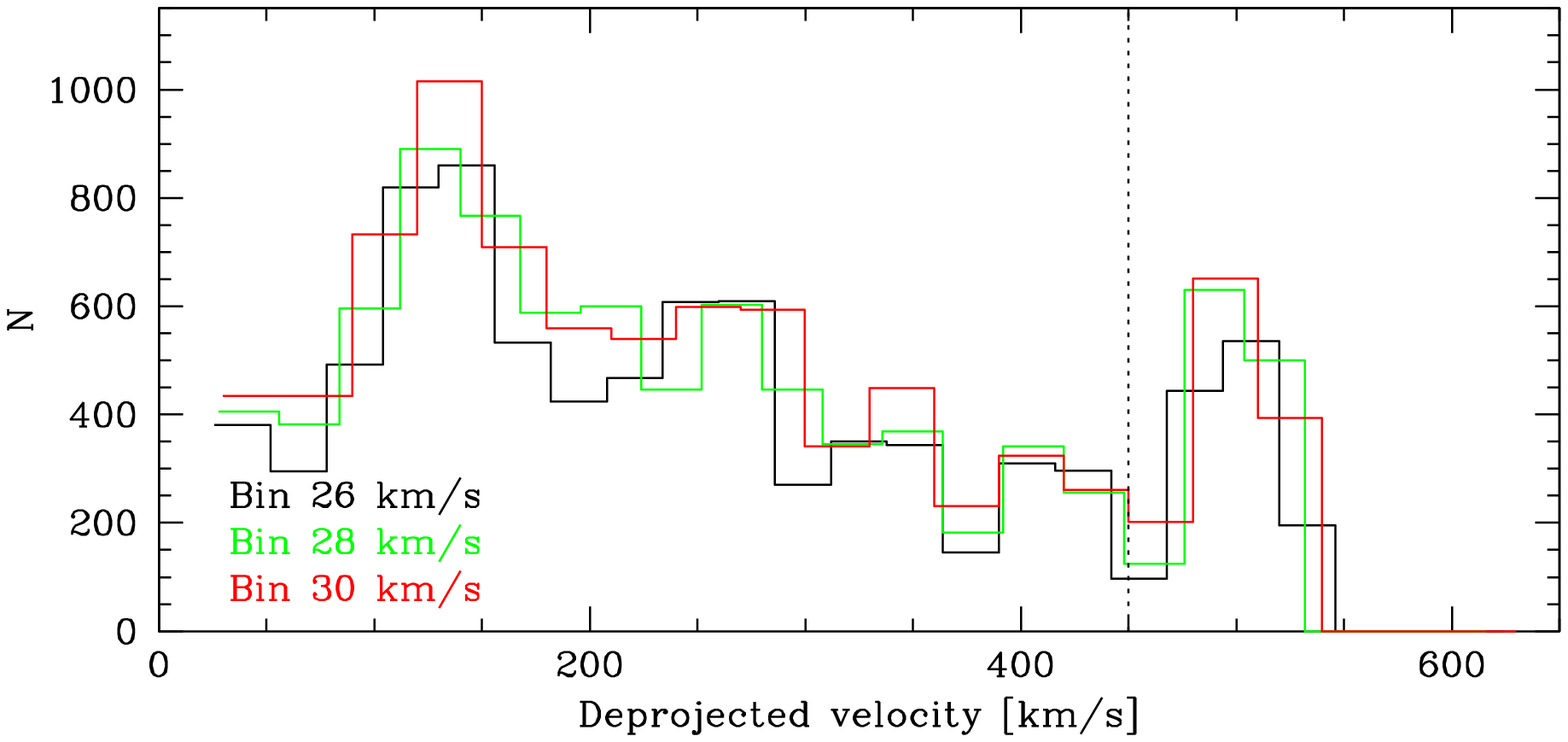}
\begin{center}
\caption{ {\bf Top panel:} de-projected 3D inter-velocity of galaxy pairs from the isolated galaxy pairs catalog (as in NC20, their Fig. 4), for the  7449 pairs with isolation parameter $\rho \geq 5$. Different bin sizes are used to show the position of the peak at $\sim 150$ km/s is stable. 
{\bf Bottom panel:} As for the top panel but considering 8571 pairs from our extended catalog. A very similar distribution is found.
In both panels, instead of limiting the plot to 450 km/s (see dotted line) as in NC20, the full range of de-projected values is shown to highlight the presence of a large number of pairs at the end of the distribution. These are pairs with 3D inter-velocity larger than the 500 km/s limit imposed to the catalog, most probably false unbound pairs. }
\label{nottaleFig4}
\end{center}
\end{figure}

The analysis of the catalog is then reported in NC20, where the statistical methods, presented and validated in Nottale \& Chamaraux (2018b, NC18b hereafter)  for de-projecting an observed radial velocity distribution, are applied.
The methods rest on the fact that the radial velocity component of a fixed 3D velocity $v$ is uniformly distributed between 0 and $v$.
NC20 applied the de-projection methods to several sub-samples of their main catalog, finding in all cases the presence of a clear excess (a peak) of preferred inter-velocities in the de-projected 3D distribution. The peak is centered at $\sim 150$ km/s and has full width half maximum FWHM $\sim 100$ km/s. This is shown in the upper panel of Fig. \ref{nottaleFig4}, which was reconstructed from the IGPC catalogue following NC20 prescriptions (their Fig. 4).
Beside the peak at $\sim 150$ km/s, the distribution of de-projected inter-velocities is approximately constant in the $0-350$ km/s range. Then,  after a local minimum around 450 km/s, a second peak  is present. This is not a real feature, rather, it is artificially produced by the de-projection technique, and represents the accumulation of those pairs with estimated 3D velocity larger than the 500 km/s limit imposed to the catalog. It is important to point out that the peak at  $\sim 150$ km/s is present, at the same location and width, in all sub-samples irrespective of the value of $\rho$ and/or filtering on  radial velocity errors (see NC20 Fig. 3,4, and 5).

Assuming most of the pairs are bound, NC20 showed that, on average, the virial mass of the pairs obtained through the 3D de-projection can be compared to their luminous mass provided the mass to light ratio is $M/L\sim30$ (in solar units). 

\begin{figure}
\centering
\hspace*{-0.5cm}
\includegraphics[scale=0.9]{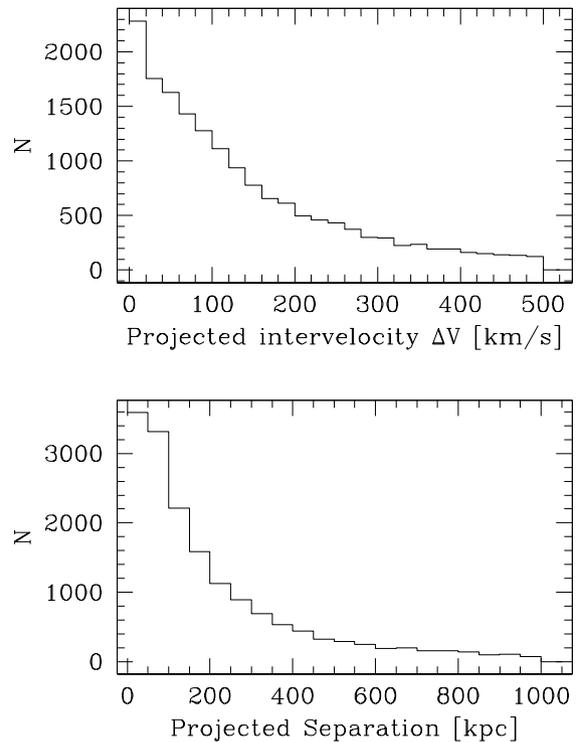}
\caption{Distribution of projected inter-velocity (upper panel) and projected separation (lower panel) for the 16404 pairs in the extend catalog. }
\label{vel_dist}
\end{figure}

\section{Extended Isolated Galaxy Pairs Catalogue}
\label{extcat}

\begin{figure}
\vspace{-0.0cm}
\centering
\hspace*{-0.5cm}
\includegraphics[scale=0.45]{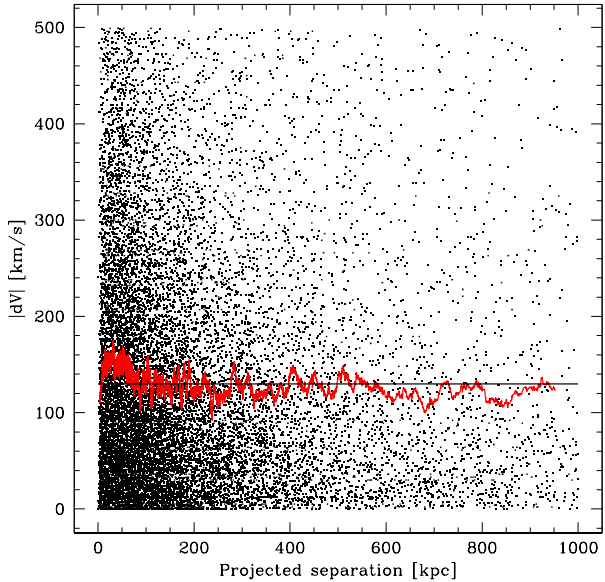}
\vspace{-1.0cm}
\caption{Observed inter-velocities versus projected separation for the 16404 galaxy pairs in the extended catalog. The red line gives the running average computed over 150 points, which is found to be sensibly constant over the whole range of projected separation. The horizontal line marks the global average of the sample.
}
\label{figDV}
\end{figure}

At the time of writing this paper the Hyperleda catalog contained data for $\sim$ 5 millions galaxies, about 20\% more than at the time NC20 built their catalog. We focus here on this new larger sample of galaxies.  There are  $\sim$ 170000 galaxies  satisfying the selection criteria stated above (see Sect \ref{Sextcat}). These were searched for pairs according to NC20 prescriptions, finding 16404 pairs with  $\rho \geq 2.5$, an increase of $\sim25$\% with respect to NC20 catalogue.
As expected and in line with NC20, most pairs are found at small projected separations and small velocity differences (see Fig. \ref{vel_dist} and \ref{figDV}). 

\begin{figure}
\centering
\hspace*{-0.6cm}
\includegraphics[scale=0.5]{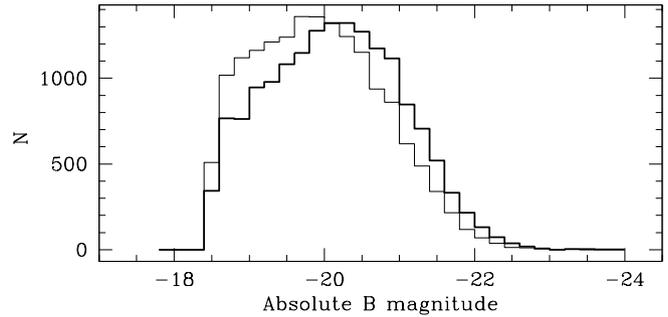}
\vspace{-5cm}
\caption{Distribution of B-band absolute magnitudes for all selected galaxies. Thick and thin histogram are for the primary and secondary component of the pairs.  }
\label{distAbsMag}
\end{figure}

\begin{figure}
\centering
\hspace*{-0.5cm}
\includegraphics[scale=0.95]{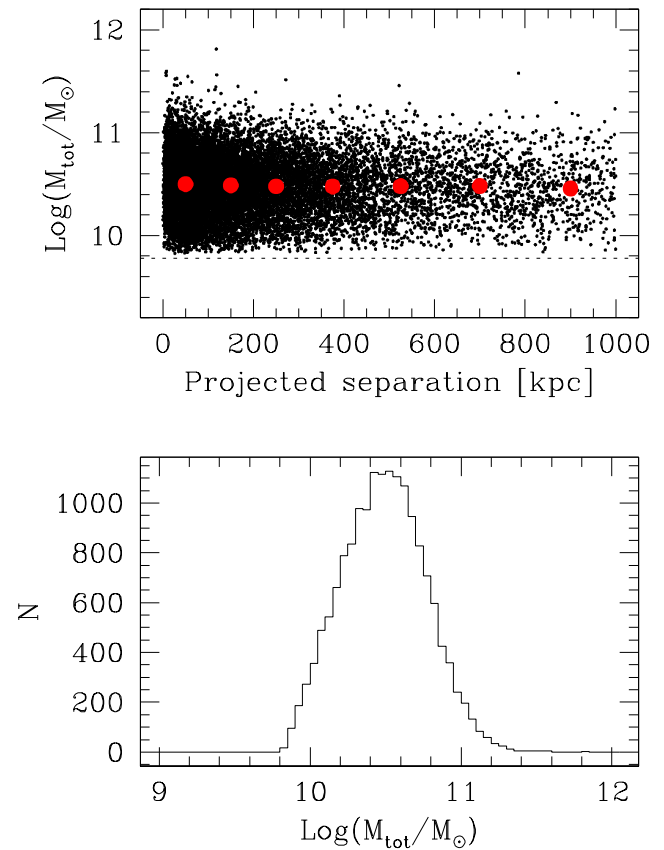}
\vspace{-0.5cm}
\caption{\textbf{Upper panel:} Total luminous mass  computed assuming M/L=1 in solar units as a function of pair separation. Red points give the average mass in bins centered at increasing projected distance. The lower mass limit (dotted line) is due to the luminosity lower limit $M(B)<-18.5$ imposed to the catalog.
\textbf{Lower panel:} Distribution of the total luminous mass showing the limited range of masses covered by the catalog. The lack of objects at the lower end of the distribution is evident. }
\label{mass_vel_dist}
\end{figure}

The absolute magnitude of the selected galaxies is restricted to a range of about 3 magnitudes being limited by the galaxy luminosity function on the bright end, and by the cutoff at  M$_B\leq -18.5$ at the faint end. The whole sample average luminosity is M$_B = 20.03$ with one $\sigma$ dispersion of 0.87 mags (Fig. \ref{distAbsMag}). The limited range of luminosity directly translates into a limited range of luminous masses, i.e. baryonic masses. Assuming M/L = 1 in solar units, masses spread a little more than an order of magnitude, with average total mass of the pairs  $3.1\times10^{10}$ M$_\odot$ (Fig. \ref{mass_vel_dist}). No dependence of the mass on the projected distance is seen.
Finally and particularly relevant for this work, applying the statistical method proposed by NC18b we derived the 3D de-projected inter-velocity for this new extended catalogue. 
The new distribution is very similar to that derived by NC20 fully confirming  the presence of a peak centered at $\sim$ 150 km/s and including about 20\% of the pairs (Fig. \ref{nottaleFig4}, bottom panel). Tests with different bins and selection criteria confirm that the position and shape of the peak is stable.

To further estimate the statistical significance of the peak, we randomly selected half of the galaxies from the 8571 pairs as used to build fig. 1. 
Two bins before and after the peak were used to set the background level. The signal in the peak was defined as the sum of three bins (centered at 100, 130, and 160 km/s, see fig. 1) after subtracting the adjacent background. 
This was repeated 1000 times, finding the peak is present  $\sim 50$\% of the times at $>5\sigma$ level, and  $\sim 90$\% of the times at $>3\sigma$ level. Note that this significance represents a lower limit  as it refers   to half of the sample.
Restricting the analysis to the 7499 pairs with accurate radial velocity ($\Delta V<$70 km/s),  the peak is even better defined being detected 60\% of the times at $>5 \sigma $ and  $\sim 97$\% of the times at $>3\sigma$ level.

\section{False Pairs Contamination}
\label{falsepairs}

A critical issue for the study of pair dynamics is to assess the level of unbound galaxy pairs present in the catalog, which was in fact built with no knowledge of whether individual pairs  were bound systems or just chance alignments. 
In the catalogue defined by  NC20 it is assumed that most of the pairs form bound systems. However, if this were the case, the Newtonian 
velocity would represent a firm upper limit and therefore, on average, the projected $\Delta V$ should decrease with the pair separation as $r^{-1/2}$.  No indication of such a behavior is observed, the average $\Delta V$ remaining  constant (Fig. \ref{figDV}). This suggests a significant fraction of pairs are unbound chance alignments. 

A rough estimate of the level of contamination can be derived assuming  the 170000 galaxies from which our extended catalog was extracted, are uniformly distributed on the sky. The corresponding  surface density is $\sim$ 4 galaxies per square degree. At a representative  radial velocity of $10000$ km/s, the adopted 1 Mpc maximum projected separation of pairs corresponds to  $\sim$ 0.5 square degrees ($H_0=70$ km/s/Mpc is used). 
Thus in each circle of radius 1 Mpc we expect to find by  chance $\sim$ 2 galaxies. 
Then, because by construction the redshift is restricted to a range of  $13000$ km/s, whatever the redshift of galaxy A, galaxy B has probability $1/13$ to have redshift within $\pm 500$ km/s. Dividing by two this probability to avoid counting pair AB and BA as two separate pairs, about 4\% or $\sim 6800$ pairs are expected by chance in our extended catalog, no matter if bound or not. Some of these candidate pairs will be eliminated by the requirement of not having other nearby galaxies within $\pm 500$ km/s. This, however, has only a minor effect because a pair is retained on average in 12 out of 13 cases. Hence, the expected level of contamination is high ($\sim$ 40\%) and probably higher  because the real data cover a significantly smaller fraction than the whole sky. 

\begin{figure}
\centering
\includegraphics[scale=0.4]{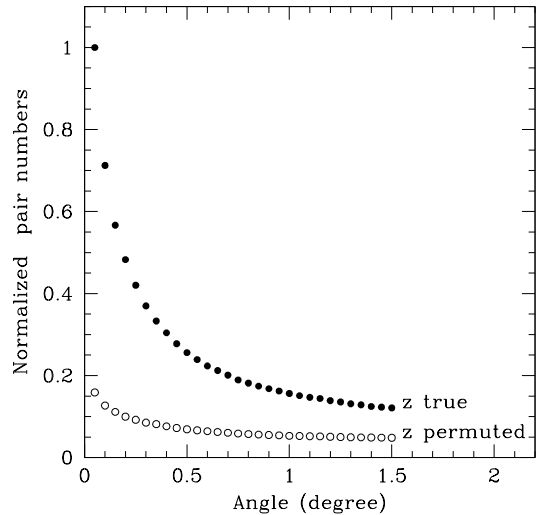}
\caption{Comparison between the number of true galaxy pairs as a function of $angular$ separation for the extended sample of $\sim$ 170000 galaxies, and the same sample after random permutation of the radial velocity.
The points are computed with a step of 0.05 degree and a redshift bin $\Delta$V = 500 km/s. 
The maximum angular separation shown (1.5 degree) corresponds to $\sim$ 1 Mpc at the lower limit of redshift in the sample.}
\label{zstest}
\end{figure}

Another way to  estimate the level of contamination is to count the number of pairs as a function of the $angular$ separation (not projected separation), and compare it to the number of pairs found after a random permutation of their redshift (see Zhdanov \& Surdej 2001 for more details). As expected the ratio between true and randomized galaxy pairs is higher at smaller separations (Fig. \ref{zstest}). 
In the explored range of angular separation (0.05 to 1.5 degree)  there is about a total factor of 3.8 between the two data-set, indicating at least 25\% of the galaxy pairs are just chance alignment of unbound pairs. 

A further estimate of the amount of contamination was obtained searching for isolated pairs in a set of galaxies constructed using the same magnitudes and positions as for the true catalog, but with randomly permuted radial velocities. The full search was performed,  including the limit in $\Delta$V, the maximum projected separation, and the isolation criterion (see Sect. 2). In this shuffled data set $\sim$ 10000 galaxy pairs were found, a value to be compared to the $\sim$ 16000 pairs found in the real catalog. Thus  $\sim$  60\% of the pairs in the extended catalogue could be unbound (just projected pairs). 

All these tests indicate that a large fraction -- possibly the majority -- of pairs in NC20 and in our extended catalogue are actually unbound  pairs.

\section{Newtonian expectations from cosmological simulations}

In order to derive the expectations on the difference of velocity between the galaxies in pairs in the context of cosmological structure formation and evolution, we refer to the Millennium Simulation (Springel et al. 2005).
In particular  Moreno et al. 2013 (hereinafter M13) used this simulation  to investigate  the dynamical behaviour of galaxy pairs embedded in a full cosmological context.
They extracted from the simulation a sample of about 1.3 million pairs having 3D separations smaller than 250h$^{-1}$ kpc. This sample includes mainly pairs formed by either one central galaxy (inside a DM halo) and a satellite galaxy, or two satellite galaxies (in a larger DM halo).
Based on these pair configurations M13 derived the expected distribution of the true velocity difference of the galaxies in the pairs and found a very broad distribution ranging from few tens to about 2000 km/s. 

No preferred orbital velocity was found by M13 in the Millenium simulation data. The distributions for the two types of pairs are shifted with respect to each other by about 400 km/s, and are both very wide having FWHM$\sim 800$ km/s, in clear contrast with the sharp peak (FWHM$\sim100$ km/s) found in our analysis of IGPC.
It is worth to note that our sample of pairs is different from that used by M13. 
First, following NC20, we defined galaxy pairs assuming a projected separation up to $1$ Mpc compared to 250h$^{-1}$ kpc
adopted in M13. Then the median mass ratio of the pairs in our sample is $\sim 1.2$, significantly smaller compared with that used by M13, that spreads over two orders of magnitude. Finally we note that while our galaxy pairs are unique and relatively isolated, those defined by M13 are not. The same galaxy in fact could  be used to define more than one pair. 

Due to all the above differences it is arduous to compare the expected inter-velocity of pairs formed in a full cosmological context to our sample. Nevertheless all M13 results clearly indicate that dynamical simulations do not predict a preferred inter-velocity for galaxy pairs. A more specific simulation better matching our galaxy selection would be important to further reinforce (or weaken) the discrepancy. 

\section{An alternative dynamical analysis}
\label{alternative}

In view of the difficulties, within the expectation of Newtonian cosmological simulations,  to explain the presence of a narrow peak in the inter-velocity distribution of galaxy pairs, 
 we explore here as an alternative  the Modified Newtonian Dynamics (Milgrom 1983a,b and c).
The basic idea of MOND is that Newtonian dynamics should be modified when the acceleration of gravity falls below a fixed value  $a_0$ .
According to this proposal, the acceleration of gravity $g$ is related to the Newtonian acceleration $g_N$ by

\begin{equation}
g_N= g \mu(g/a_0).
\end{equation}

The interpolation function $\mu(g/a_0)$, which is not defined by the theory, admits the asymptotic behavior $\mu(g/a_0)=1$ for $g>>a_0$, so to retrieve the Newtonian expression in the strong field regime, and $\mu(g/a_0)=g/a_0$ for $g<<a_0$.

The value of $a_0$ must be derived observationally.  Studying local galaxies Begeman, Broeils, and Sanders (1991)  found $a_0 \sim 1.2\times 10^{-8}$ cm/$s^2$, which is the most widely adopted value. Later studies, however, proposed different values from as low as 0.9 (Bottema et al. 2002) to as high as 1.4 (see discussion in Gentile et al. 2011).

Since the seminal papers by Milgrom (1983a,b,c), MOND has been applied to several astrophysical objects including (in order of increasing size) 
wide binary stars (Hernandez, Jiménez \& Allen 2012; Hernandez, Cookson \& Cortes 2021, Scarpa et al 2017), Globular clusters (Scarpa, Marconi \& Gilmozzi 2003; Scarpa \& Falomo 2010; Scarpa et al. 2011; Hernandez \& Lara-D I  2020), 
dwarf galaxies (Milgrom 1995; McGaugh \& Milgrom 2013; Sanders 2021), 
gas dominated galaxies (McGaugh 2012; Sanders 2019),  
spiral galaxies (Sanders 1996; Gentile, Famaey \& de Blok 2011; Milgrom \& Sanders 2007), elliptical galaxies (Milgrom \& Sanders 2003; Durazo et al. 2018; Tian \& Ko 2016), pair of galaxies (Milgrom 1983c),  group of galaxies (Milgrom 2019; McGaugh et al. 2021), gravitational lenses (Sanders 2014), and 
cluster of galaxies (Sanders 1999, 2003). In all but one case MOND succeeds in describing the observations without the need of DM. The only case in which an excess of mass (by a factor two) is found between observations and MOND prediction is in rich clusters of galaxies. This is a long standing, unsolved problem for MOND that, however, does not falsify the theory because it points to the existence of still to be discovered baryonic mass in clusters. 
For further reading and recent reviews see Famaey \& McGaugh (2012), and Merritt (2020).

Assuming there are only baryons, the mass distribution in galaxies is such that the MOND regime is reached for distances smaller or comparable to the size of the galaxies themselves. Thus, when considering the dynamics of galaxy pairs, even the closest ones, the exact shape of the interpolation function $\mu$ is irrelevant. Everything takes place in deep MOND regime where the acceleration becomes 
\begin{equation}
g= \sqrt{ g_N a_0}.
\end{equation}

At present, no fully fledged numerical simulations developed within the framework of MOND exist. Therefore, only a crude approximation can be made of what galaxy pair dynamics should be. Here we consider the approximate formula eq. 3 proposed by Milgrom (1983c) giving the expected velocity difference $\Delta V$  for galaxy pairs in the simple case of circular motion and equal mass M of the two components:

\begin{equation}
\Delta V^4= 2 G M_{tot} a_0
\end{equation}

where $M_{tot} = 2M$ is the total mass of the pair. Note the lack of dependence of $\Delta V$ on the pair  separation. This is a key characteristic of MOND which recovers the Tully-Fisher relation (Tully \& Fisher 1977). 

\begin{figure}
\centering
\hspace*{-1.0cm}
\includegraphics[scale=0.55]{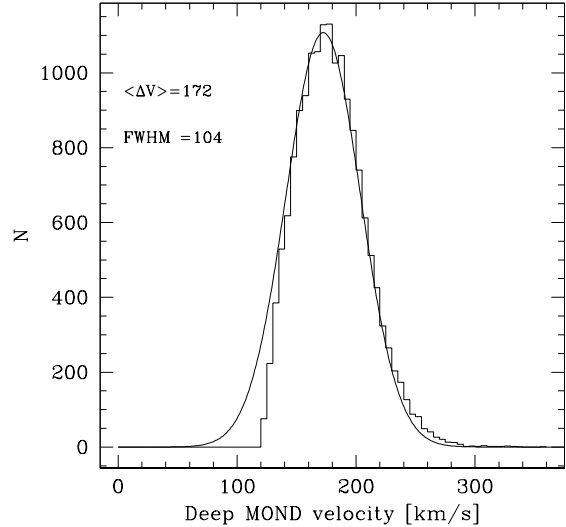}
\caption{Predicted deep MOND intervelocity distribution, computed according to eq. 3 and assuming a mass to light ratio M/L=1 (in solar units). Data are well described by a Gaussian (with average and full width at half maximum as indicated). The mismatch at the lower velocities being due to the missing low mass pairs. }
\label{deep_mond_vel}
\end{figure}


In order to reconstruct the distribution of the orbital velocity under the MOND framework one should start from the luminous mass distribution (Fig. \ref{mass_vel_dist}) using the formula (3). The resulting inter-velocity distribution is well matched by a Gaussian with average $<V>=172$ km/s and FWHM $= 104$ km/s (Fig. \ref{deep_mond_vel}).
The lack of dependence of the velocity on separation, and the
$M^{1/4}$ dependence of the velocity in MOND, further combined with the luminosity limits applied to the catalog, has the effect that only a very limited range of orbital velocities for galaxy pairs is permitted. Neither too small nor too large velocities are allowed. 
The MOND distribution clearly suggests the existence of a region of preferred velocity, which directly compare to the peak at $\sim$ 150 km/s observed in the 3D de-projection of the inter-velocity 
(Fig. \ref{nottaleFig4}). 
Both peaks have similar width, though are shifted by about $\sim$ 20 km/s.
In this framework therefore the pairs outside the peak represents unbound systems.

\section{Discussion and Conclusions}
\label{discusion}

We  analyzed the properties of the 3D inter-velocity distribution of a sample of isolated galaxy pairs.
The sample is an important laboratory for studying  galaxy pair dynamics and  led to the remarkable discovery of a preferred 3D inter-velocity of $\sim 150$ km/s (NC20).
Within the framework of Newtonian dynamics, there is no straightforward explanation for the existence of a preferred orbital velocity, a point already recognized by NC20. 
While an explanation in the  Newtonian dynamics with some {\it ad hoc } hypothesis (e.g., an adequate dark matter distribution)  cannot be excluded, it is of interest to explore alternative interpretations. 
In particular, for this sample of galaxy pairs MOND predicts the existence of a narrow distribution of orbital velocities --as wide as the observed one -- centered at ~170 Km/s. The limited range of orbital velocities is due to  the pairs being in deep MOND regime, so that the pair separation becomes irrelevant in determining the orbital velocity. 
The observed peak accounts for $\sim 20$ \% of the total galaxy pairs. In this scenario the remaining pairs represent unbound systems as discussed 
 in Sect. \ref{falsepairs}.

Note that the peak expected in the  MOND interpretation is at $\sim$ 170 km/s while that derived from the de-projection analysis is at $\sim$ 150 km/s. Taking at face value this 20 km/s difference, to match the observations to the prediction from eq. 3 -- which is itself approximate (Milgrom. 1983c) -- a reduction of a factor $\sim 2$ is required on the right-hand side of eq. 3. 
Among the many possible sources for this mismatch we mention: 
i) $a_0$ could be smaller than the generally adopted value, 
ii) the bound pairs might have on average lower luminosity than the unbound pairs, 
iii) the mass to light ratio for bound systems could be significantly different than the value adopted here. 
For instance taking $a_0=0.9\times 10^{-8}$, a value at the lower end of the one reported in the literature (Bottema et al. 2002), the mismatch is halved (peak center at 160 km/s). 
Further computing luminous masses adopting H$_0=74$ km/s/Mpc which is more appropriate for the local Universe (Riess et al. 2019), instead of 70 km/s/Mpc as used in Hyperleda, would additionally move the peak velocity to 155 km/s. 
On the other hand, if the mass to light ratio were significantly higher than  M/L=1 adopted here the mismatch would be larger. For M/L=2 the peak velocity would move to $\sim$ 200 km/s, a value uncomfortably large for the MOND interpretation. 

It is clear, however, that the relevant aspect of this investigation is not the exact position of the peak, which  depends on a number of parameters and the adopted modification of Newtonian gravity. 
Rather, it is that the velocity distribution of isolated galaxy pairs may be relevant for constraining the dynamics in the low acceleration regime, and in this contest alternative dynamical scenarios, like 
MOND, should be taken in serious consideration.

\vspace{1cm}

{\bf Data Availability:} The data supporting the findings of this study are openly available at Hyperleda web site (http://leda.univ-lyon1.fr).
\vspace{1cm}

\section*{Acknowledgments}
We are grateful to the anonymous  referee for useful comments and suggestions.

{\bf REFERENCES} \\
\vspace{1cm}

   Begeman K. G., Broeils A.H. ans Sanders R. H. 1991, MNRAS 249, 523
\\    Bottema R., Pestaña J. L. G., Rothberg B. \& Sanders R. H. 2002, A\&A, 393,453
\\    Durazo R., Hernandez X., Cervantes Sodi B. \& Sanchez S. F. 2018, ApJ 863, 107
\\    Famaey B. and McGaugh S. S. 2012, LRR 15, 10
\\    Gentile G.,  Famaey B., \& de Blok W. J. G. 2011, A\&A 527, 76
\\    Hernandez X. \& Lara-D I A. J. 2020, MNRAS 491, 272
\\    Hernandez X., Jim\'enez M. A. \& Allen, C. 2012, EPJC 72, 1884 
\\    Hernandez X., Cookson S., Cortes R. A. M. 2021, arXiv:2107.14797
\\    Makarov D., Prugniel P., Terekhova N., Courtois, H. \& Vauglin I. 2014, A\&A, 570, 13
\\    McGaugh S. S. 2012 AJ 143, 40
\\    McGaugh S. S. \& de Blok, W. J. G. 1998, ApJ 499, 66.
\\    McGaugh S. S. \& Milgrom M. 2013, ApJ 766, 22
\\    McGaugh S. S., Lelli F., Schombert J. M. et al. 2021, AJ 162, 202
\\    Merritt D. 2020, "A philosophical approach to MOND",  Cambridge University Press
\\    Milgrom M. 1983a, ApJ 270, 365 
\\    ------. 1983b, ApJ 270, 371
\\    ------. 1983c, ApJ 270, 384
\\    ------. 1994, ApJ 429, 540
\\    ------. 1995, ApJ 455, 439
\\    ------. 2019, PhRvD 99, 4041
\\    Milgrom M. \& Sanders R. H. 2003, ApJL 599, 25 
\\    ------. 2007, ApJL 658, 17 
\\    Nottale L. and Chamaraux P. 2018, Astrophysical Bulletin, 73, 310
\\    Nottale L. and Chamaraux P. 2018b, A\&A, 614, 45 
\\    Nottale L. and Chamaraux P. 2020, A\&A 641, 115
\\    Sanders R. H. 1996, ApJ 473, 117
\\    ------. 1999, ApJL 512, 23
\\    ------. 2003, MNRAS 342, 901
\\    ------. 2014, MNRAS 439, 1781
\\    ------. 2019, MNRAS 485, 513
\\    ------. 2021, MNRAS 507, 803
\\    Riess A. G., Casertano S., Yuan W. et al. 2019, ApJ 876, 85
\\    Scarpa R. \& Falomo 2010, A\&A 523, 43  
\\    Scarpa R., Marconi G. \& Gilmozzi R. 2003, A\&AL 405, 15
\\    Scarpa R. Marconi G. Carraro G. et al. 2011, A\&A 525, 148
\\    Scarpa R., Ottolina R., Falomo R., Treves A., 2017, IJMPD, 26, 1750067. doi:10.1142/S0218271817500675
\\    Springel V., et al., 2005, Nat, 435, 629
\\    Tian Y \& Ko C. 2016, MNRAS 462, 1092
\\    Tully R. B. and Fisher J. R. 1977, A\&A 54, 661 
\\       Zhdanov V. I. and Surdej J. 2001, A\&A, 372, 1

\end{document}